\documentclass{article}

\usepackage{graphicx}
\usepackage{multirow}
\usepackage{caption}
\usepackage{amssymb, amsmath}
\usepackage{url}
\usepackage{authblk}
\usepackage{multirow}

\begin{document}
\title{Development of an Estimation Method for the Seismic Motion Reproducibility of a Three-dimensional Ground Structure Model by combining Surface-observed Seismic Motion and Three-dimensional Seismic Motion Analysis}
\author[1]{Tsuyoshi Ichimura}
\author[1]{Kohei Fujita}
\author[1]{Ryota Kusakabe}
\author[2]{Hiroyuki Fujiwara}
\author[3]{Muneo Hori}
\author[1]{Maddegedara Lalith}
\affil[1]{Earthquake Research Institute and Department of Civil Engineering, The University of Tokyo, Japan}
\affil[2]{Multi-hazard Risk Assessment Research Division, National Research Institute for Earth Science and Disaster Resilience, Japan}
\affil[3]{Research Institute for Value-Added-Information Generation, Japan Agency for Marine-Earth Science and Technology, Japan}
\date{}
\maketitle

\begin{abstract}
The ground structure can substantially influence seismic ground motion underscoring the need to develop a ground structure model with sufficient reliability in terms of ground motion estimation for earthquake damage mitigation.
While many methods for generating ground structure models have been proposed and used in practice, there remains room for enhancing their reliability.
In this study, amid many candidate 3D ground structure models generated from geotechnical engineering knowledge, we propose a method for selecting a credible 3D ground structure model capable of reproducing observed earthquake ground motion, utilizing seismic ground motion data solely observed at the ground surface and employing 3D seismic ground motion analysis.
Through a numerical experiment, we illustrate the efficacy of this approach.
By conducting $10^2$--$10^3$ cases of fast 3D seismic wave propagation analyses using graphic processing units (GPUs), we demonstrate that a credible 3D ground structure model is selected according to the quantity of seismic motion information.
We show the effectiveness of the proposed method by showing that the accuracy of seismic motions using ground structure models that were selected from the pool of candidate models is higher than that using ground structure models that were not selected from the pool of candidate models.
\end{abstract}

\section{Introduction}

Structures on or within the ground can experience substantial impacts from the ground structure during seismic events \cite{review}\cite{tohokueqreview}.
For instance, soft ground atop the hard ground, coupled with the characteristics of seismic motion and the ground structure, can lead to localized amplification of seismic motion, resulting in damage.
Enhancing the reliability of ground structure assessments is required to enable accurate evaluation of structural behavior during seismic events and mitigate earthquake-induced damage effectively.

Many methods for estimating ground structures have been proposed and employed in practical applications.
For instance, methods directly assessing {\it in situ} geotechnical properties, such as borehole testing, yield reliable outcomes.
Nonetheless, due to the limited number of measurement points, interpolation becomes necessary to estimate the spatial extent of the ground structure.
In the construction of 3D ground structure models, interpolation methods based on geotechnical engineering knowledge, such as inverse distance weighting methods \cite{IDW}, curvature minimization principles \cite{mincurve}, and Kriging methods \cite{kuriging}, are employed.
These interpolations rely on various assumptions, resulting in multiple 3D ground structure models, necessitating additional evaluation to determine the reliability of each model.

Such evaluation can utilize the observed ground motion of small amplitudes.
For instance, the reliability of a candidate ground structure model can be assessed by comparing the phase velocity derived from long-term microtremor observations \cite{microtremor} or by comparing the Green's functions obtained via seismic interferometry \cite{kansho}.
Alternatively, the quality of 3D ground structure models can be evaluated by directly utilizing observed seismic ground motions of small amplitudes, which typically exhibit stronger signal strengths.
For example, \cite{yamaguchiiccs} demonstrates the feasibility of evaluating a 3D ground structure model using observed seismic ground motions of small amplitudes.
However, the method in \cite{yamaguchiiccs} assumes that seismic motions input to the 3D ground structure model are observed at underground observation points, making it challenging to apply to sites where underground observation points cannot be installed.

Building upon the above insights, we propose a method that selects a credible ground structure model from many generated 3D ground structure models using small amplitude seismic motions solely observed at the ground surface.
We illustrate the effectiveness of the method through numerical experiments
by conducting $10^2$--$10^3$ cases of fast 3D seismic wave propagation analysis on GPUs, and show that we can select a credible ground structure model based on the amount of seismic motion information observed at the ground surface.
Additionally, we show that the selected 3D ground structure model can be utilized to evaluate ground motion with sufficient accuracy.

\section{Method}

\begin{figure}[tb]
\includegraphics[width=\hsize]{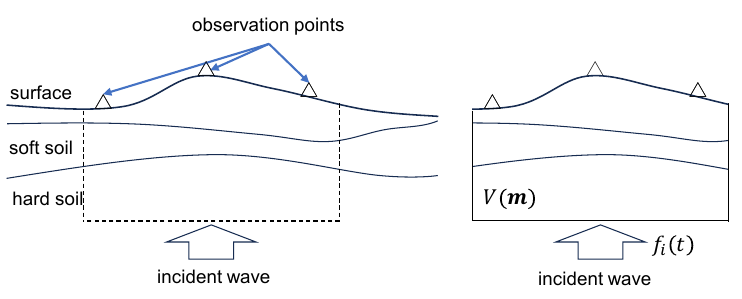}
\caption{Target system (left) and its numerical analysis model (right).}
\label{fig:problemsetting}
\end{figure}

We introduce a method for extracting a credible 3D ground structure model from a pool of candidate 3D ground structure models using small amplitude seismic motions observed at the ground surface.
The general setting, depicted on the left of Fig.~\ref{fig:problemsetting}, involves inputting seismic waves to a 3D ground structure comprising soft and hard soil, with resulting seismic ground motions observed at the designated points on the surface.
Notably, the 3D ground structure and the input seismic waves are typically unknown.

Next, we elucidate the numerical analysis model for the target system, as depicted on the right side of Fig.~\ref{fig:problemsetting}.
This study uses the cartesian coordinate system $x_1, x_2, x_3$ for simplicity.
The ground enclosed by the dashed line in the left panel of Fig.~\ref{fig:problemsetting} is designated as $V(\mathbf{m})$, defined by a set of model parameters $\mathbf{m}$.
An external force, represented by the input velocity wave $f_i(t)$, is applied to the bottom of $V(\mathbf{m})$, and its dynamic response is analyzed to yield the time-history response $obs^k_i(t)$ at each observation point $k$ on the ground surface.
To compute the small amplitude seismic response of the ground to an earthquake, we model $V(\mathbf{m})$ as linearly elastic and solve the governing equations for a linear dynamic elastic body
\begin{equation}
\label{GE:ORG}
d_i (c_{ijkl},d_k u_l)=\rho \ddot{u}_j.
\end{equation}
Here, $c_{ijkl}$, $u_j$, and $\rho$ represent the elasticity tensor, displacement in the $j$-th direction, and density, respectively.
The notation $(\ddot{~})$ denotes second-order derivatives in time, and $d_i$ indicates the differentiation in the $i$-th direction.
Note that stress-free, non-reflective, and semi-infinite absorbing boundary conditions are enforced on the model's top, bottom, and side surfaces.
We aim to develop a method that can evaluate $V(\mathbf{m})$ and $f_i(t)$ that is consistent with $obs^k_i(t)$.

We describe the parametrization employed for this evaluation.
First, we suppose that the input velocity wave can be described as follows:
\begin{equation}
f_i(t)=\sum_j c_{ij} p(t-(j-1) \Delta t). \nonumber
\end{equation}
We utilize unit impulse waves $p(t)$, for instance, trigonometric functions, to represent the input velocity wave.
Here, $c_{ij}$ represents unknown scalar values assessed through error minimization, and $\Delta t$ denotes the time-stepping stride for discretization.
While arbitrary $p(t)$ and $\Delta t$ are permissible, we configure them to facilitate the reconstruction of the input wave within the target frequency range.
Moreover, we denote the Green's function of the $i$-th directional velocity response at observation point $k$ when $p(t)$ is input in the $j$-th direction at the bottom of $V(\mathbf{m})$, as $G^k_{ij}(\mathbf{m},t)$.
The $i$-th directional velocity at observation point $k$ for $f_j(t)$ can thus be expressed as
\begin{equation}
U^k_i(\mathbf{m}, t)=\sum_{j=1}^3 \sum_l G_{ij}^k(\mathbf{m}, t-(l-1)\Delta t) c_{jl}. \nonumber
\end{equation}
Utilizing the observed ground velocity in the $i$-th direction at the $k$-th observation point ($obs^k_i(t)$), the error can be assessed as follows:
\begin{equation}
ERR(\mathbf{m})=\frac{1}{3 n_{obs}}\sum_{k=1}^{n_{obs}} \sum_{i=1}^3  \frac
{\sqrt{ \int (U^k_i(\mathbf{m}, t)-obs^k_i(t))^2 \mathrm{d}t}}
{\sqrt{\int obs^k_i(t)^2 \mathrm{d}t}}.
\label{eqerr}
\end{equation}

With the parameter settings outlined above, we compute $c_{ij}$ that minimizes the error from the observed seismic motion $obs^k_i(t)$ for each candidate $l$-th 3D ground structure model characterized by model parameters $\mathbf{m}_l$.
First, we compute $G^k_{ij}(\mathbf{m}_l,t)$ for the model parameters $\mathbf{m}_l$.
Subsequently, by considering the stationary condition of $ERR$, which is a second-order function of $c_{ij}$, 
the coefficients $c_{ij}$ can be determined by solving:
\begin{equation}
\mathbf{A} \mathbf{c}= \mathbf{b}. \nonumber
\end{equation}
Here, $\mathbf{A}$, $\mathbf{b}$ represent a constant matrix and a constant vector, respectively, while $\mathbf{c}$ denotes an unknown vector with components $c_{ij}$.

By applying singular value decomposition to $\mathbf{A}$ and discarding negligible singular values, we construct a pseudoinverse matrix and compute $\mathbf{c}$ and $ERR$. Note that the dimension of $\mathbf{A}$ is small because this computation can be performed independently for each trial, and thus the cost of this computation can be kept small enough.
That is, if the seismic ground motions contain sufficient information (i.e., the number of observation points and the number of events is adequate) and are suitably constrained by the model-specific time-history Green's functions, it is possible to assess the credibility of various 3D ground structure models by comparing the error $ERR$ obtained through attempts to reproduce observed seismic ground motions using $\mathbf{m}_l$.

\section{Numerical Experiment}

We conducted a numerical experiment to demonstrate the efficacy of the proposed method in selecting a credible 3D ground structure model from among geotechnically estimated ground models, using seismic ground motions observed at the ground surface.
Here, a credible ground model can reproduce the seismic ground motions observed at the surface.
Specifically, the method described in Section 2 estimates the seismic ground motion that best fits the observed seismic ground motion using a given 3D ground structure model, albeit with introduced errors.
This analysis in estimating seismic ground motion that best fits the observed ground motion for a given 3D ground structure model is conducted each time an earthquake is observed.
Discarding ground models that inadequately reproduce seismic ground motions can identify a credible ground model capable of reproducing observed seismic ground motions at a specific site.

\begin{figure}[tb]
\centering
\includegraphics[width=0.9\hsize]{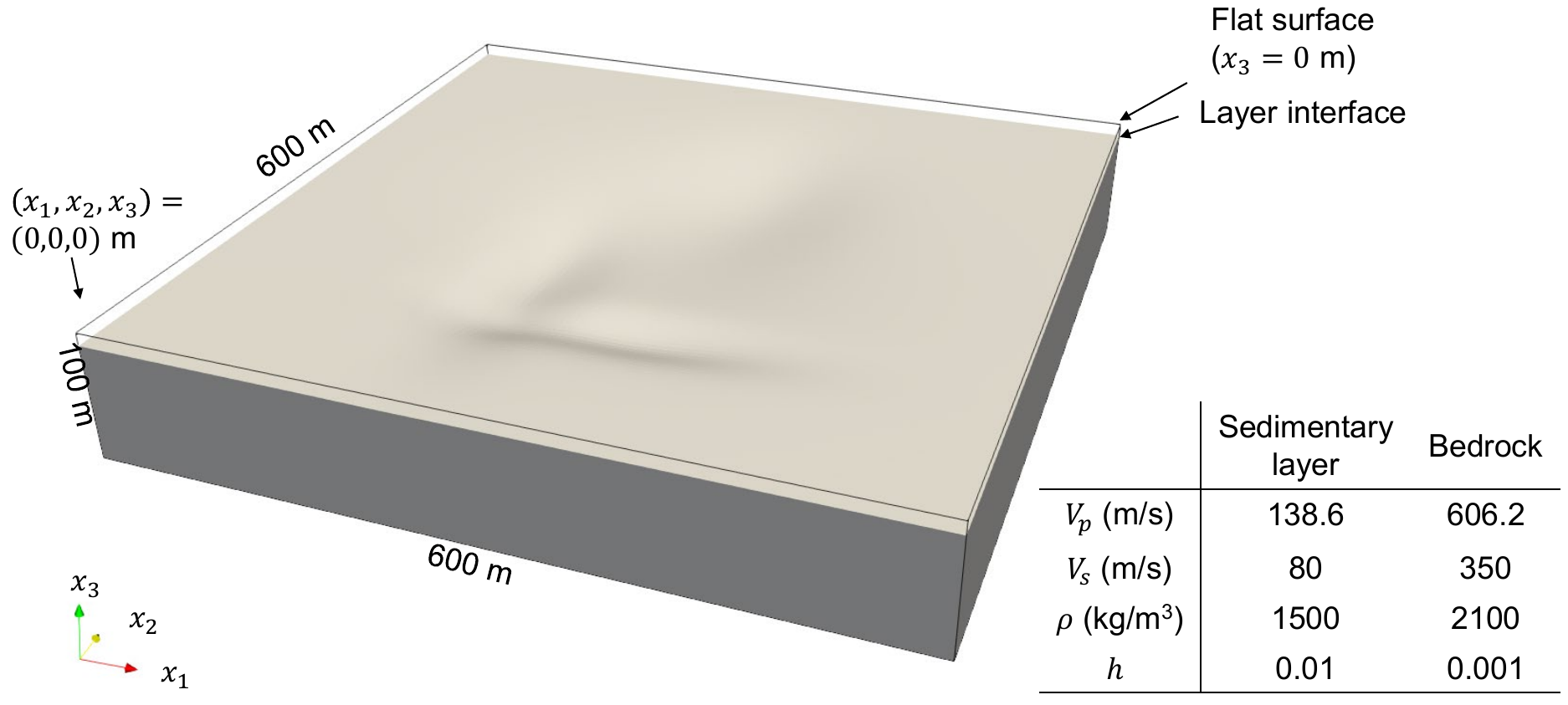}
\caption{Reference ground model. The model comprises two layers, with a flat surface and a sedimentary layer with varying thickness. The thickness of the sedimentary layer is illustrated in Fig.~\ref{fig:dem}a).}
\label{fig:refmodel}
\end{figure}

\begin{figure}[tb]
\centering
\begin{minipage}[c]{0.32\hsize}
\centering
\includegraphics[height=3.5cm]{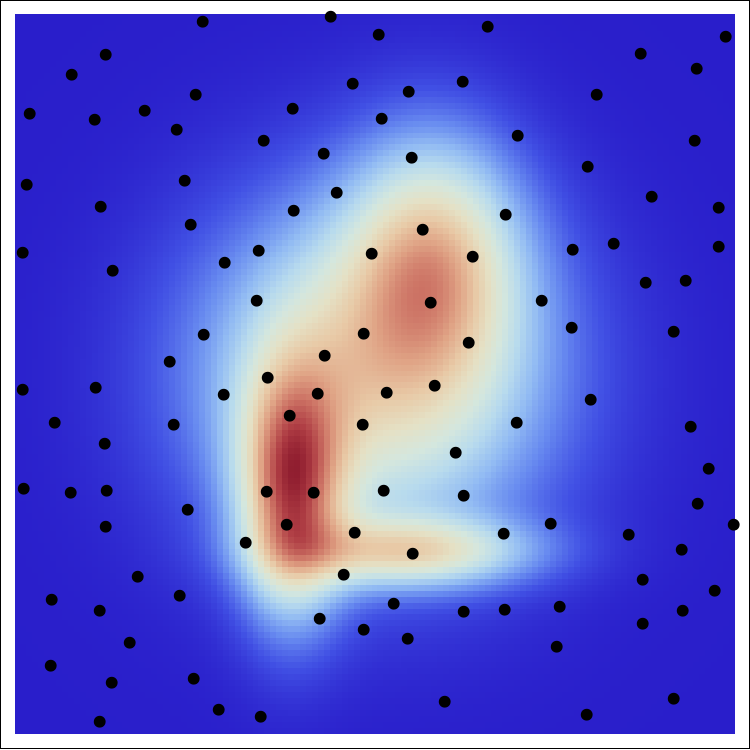}\\
a) Reference model
\end{minipage}
\begin{minipage}[c]{0.32\hsize}
\centering
\includegraphics[height=3.5cm]{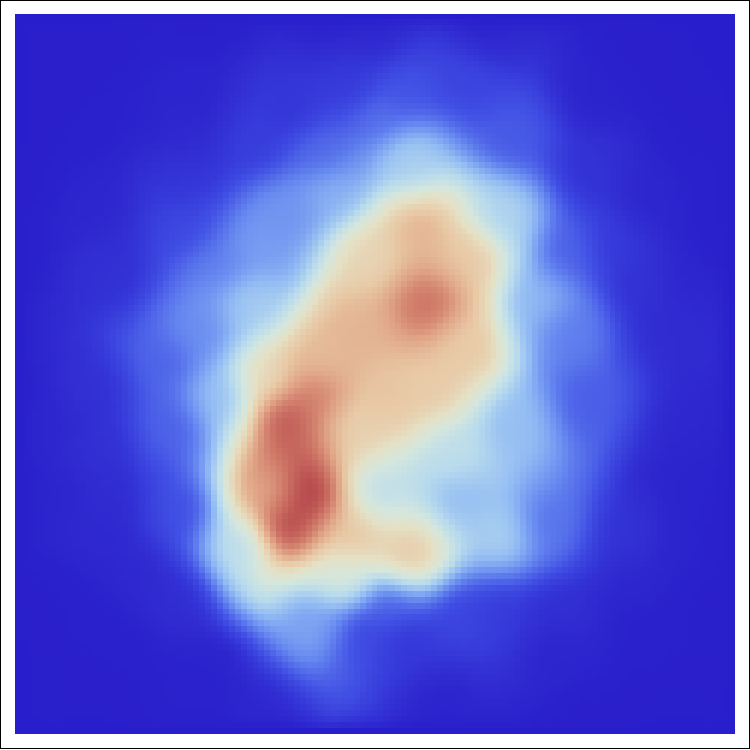}\\
b) Model000018
\end{minipage}
\begin{minipage}[c]{0.32\hsize}
\centering
\includegraphics[height=3.5cm]{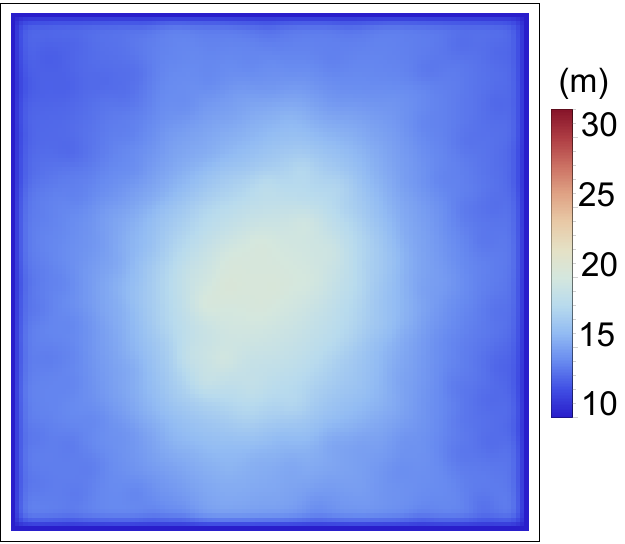}\\
c) Model000228
\end{minipage}
\caption{Thickness of sedimentary layers for each model. The position of 120 boring log points is also indicated in a).} 
\label{fig:dem}
\end{figure}

First, we set a reference model to serve as the correct solution for the numerical experiment.
The reference model comprises a two-layer structure comprising a sedimentary layer and bedrock, which imitates a real ground structure.
It spans 600, 600, and 100 m in the $x_1$, $x_2$, and $x_3$ directions, respectively (see Fig.~\ref{fig:refmodel}).
The ground surface is flat at an elevation of 0 m, with the thickness of the first layer specified as depicted in Fig.~\ref{fig:dem}a).
The physical properties of each layer are given in Fig.~\ref{fig:refmodel}. 
This ground structure emulates valley floor lowlands formed by sediment accumulation from river meandering.
During earthquakes, local ground amplification occurs, emphasizing the importance of accurately assessing the ground structure for evaluating seismic ground motion and making effective earthquake mitigation strategies.
Assuming that the reference model is situated at KiK-net \cite{KiK-net} station IBRH19, we simulate a scenario where earthquakes listed in Table~\ref{eqtable} occur sequentially, and seismic motions are observed.
To simulate this scenario, actual ground motions observed at underground stations of IBRH19 were inputted as forces from the bottom of the reference ground model, and pseudo-observed seismic ground motions were recorded at the ground surface.
The analysis was performed up to 2.5 Hz, which is the frequency range considered to have a significant impact on structural damage.
Semi-infinite absorbing boundary conditions are applied to the sides and bottom of the numerical model.

\begin{table}[tb]
\centering
\caption{Properties of the earthquakes used in the numerical experiment}
\label{eqtable}
\begin{tabular}{ccccc}
Event \#  & Data time (JST) & Epicenter & Depth (km) & Magnitude \\ \hline
1 & 2023/02/25 22:27 & 42.755N 145.075E & 63 & 6.0 \\
2 & 2023/03/27 00:04 & 38.307N 141.615E & 60 & 5.3 \\
3 & 2023/05/11 04:16 & 35.170N 140.185E & 40 & 5.2 \\
4 & 2023/07/29 19:34 & 36.347N 139.958E & 77 & 4.6 \\
5 & 2023/12/22 01:11 & 35.238N 141.137E & 10 & 4.8 \\
6 & 2024/01/28 08:59 & 35.6N 140.0E & 80 & 4.8 \\ \hline
  \end{tabular}
\end{table}

Next, a set of candidate ground structure models is generated using geotechnical engineering methods.
While many methods generating ground structure models exist, we assume that the physical properties of the first and second layers, along with the boundary location between them, are determined from a simple ground survey conducted at the 120 survey points illustrated in Fig.~\ref{fig:dem}a) (i.e., $\mathbf{m}$ represents model parameters describing the boundary shape between the first and second layers).
The geometry of the boundary between the first and second layers within the target area of 600 $\times$ 600 m is estimated based on information regarding the boundary layer location at these 120 survey points, creating a set of candidate ground structure models.
Although many methods for estimating the layer boundary geometry under these conditions exist, we employ an inverse distance weighting method.
This method assigns weights to observed values obtained near the evaluation point based on the inverse of the distance, yielding an interpolated result.
Utilizing the inverse distance weighting method, the layer thickness $z$ at any point in the domain is evaluated as follows:
\begin{equation}
z(x_1,x_2)=\sum_{i=1}^M \frac{\bar{z}_i/d_i^q}{\sum_{i=1}^M 1/d_i^q},~{\rm where}~d_i=\sqrt{(x_1-\bar{x}_1^i)^2+(x_2-\bar{x}_2^i)^2}, \nonumber
\end{equation}
where $\bar{z}_i$ denotes the layer thickness at the $i$-th measurement point located at $(\bar{x}_1^i, \bar{x}_2^i)$, $M$ represents the number of measurement points nearest to the target position $(x_1,x_2)$ utilized for interpolation, and $q$ is a positive constant parameter.
The values of $M$ and $q$ are arbitrary.
In this context, we designate $M=i ~(i=1,2,...,50)$ and $q=0.1 i ~(i=1,2,.... ,40)$ to generate a total of $50 \times 40 =2000$ 3D ground structure models.
Subsequently, from the pool of candidate 3D ground structure models, we select 236 models that exhibit substantial variations in the layer boundary geometry, based on a 3D ground structure model generated using commonly used parameters (i.e., $M=20$ and $q=2.0$).
Note that Laplace smoothing was applied five times to mitigate steep slopes in certain areas.

\begin{figure}[tb]
\centering
\includegraphics[height=4.2cm]{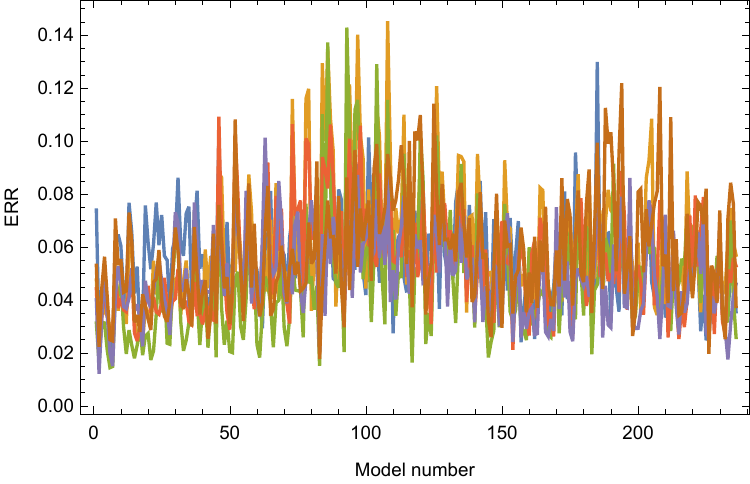}
\includegraphics[height=4.2cm]{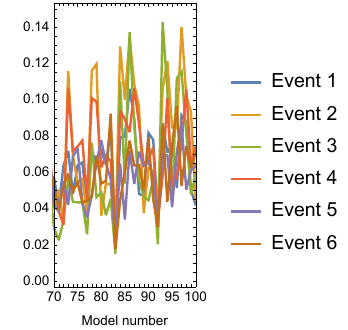}\\
a) $ERR$ for each event (left: overview, right: closeup) \\
\includegraphics[height=4.2cm]{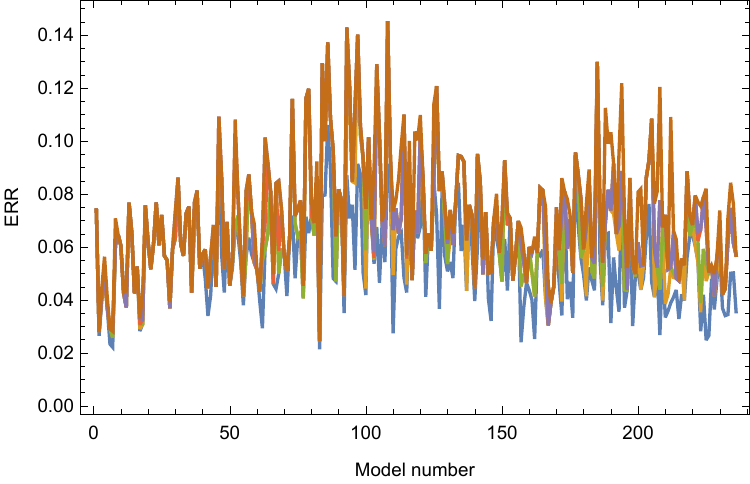}
\includegraphics[height=4.2cm]{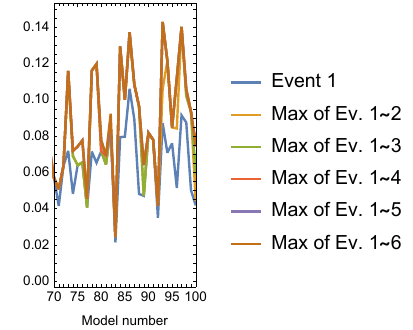}\\
b) Maximum $ERR$ for events (left: overview, right: closeup) \\
\caption{Estimated $ERR$ using one observation point} 
\label{fig:errornobs1}
\end{figure}

\begin{figure}[tb]
\centering
\includegraphics[width=0.8\hsize]{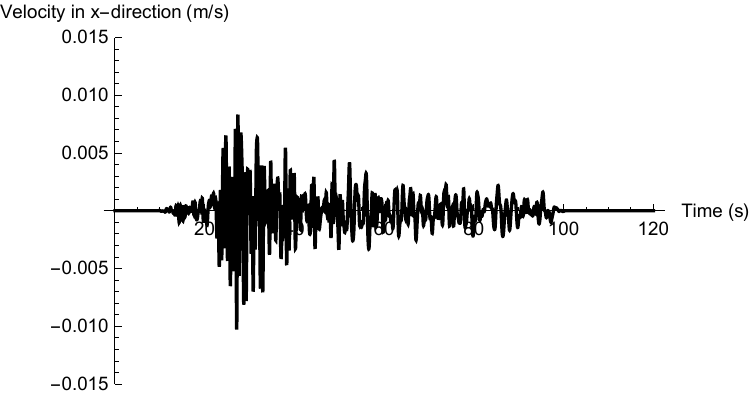}\\
a) True incident wave \\
\includegraphics[width=0.8\hsize]{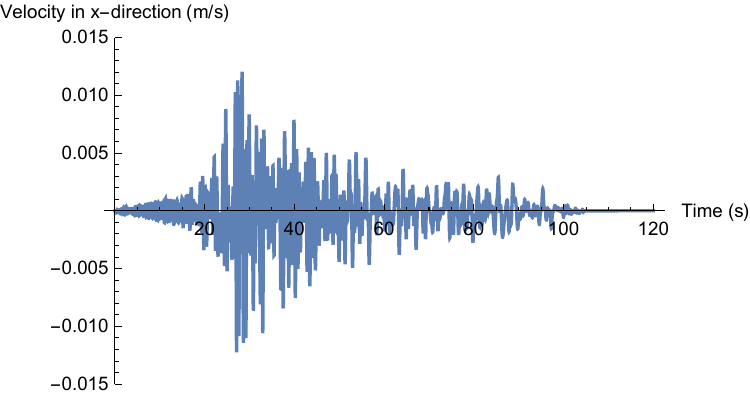}\\
b) Estimated incident wave using model000083 ($ERR=0.0243$) \\
\caption{Estimated incident wave using one observation point for event \#1. Estimation accuracy is low even if $ERR$ is low.}
\label{fig:inputwavenobs1}
\end{figure}

Utilizing the method described in Section 2 and the pseudo-observed seismic ground motions described above, we attempt to identify credible 3D ground structure models by reproducing the observed seismic ground motions for each of the 236 obtained models.
First, we consider a scenario where only one observation point is available on the ground surface at $(x_1, x_2)=(300, 300)$ m.
The analysis conducted for each candidate model is the same as the analysis using the reference model (i.e., the target frequency is set up to 2.5 Hz), with semi-infinite absorbing boundary conditions applied to the sides and bottom of the model (the same applies to the other analysis cases in this Section).
Figure~\ref{fig:errornobs1}a) illustrates the errors obtained for each ground structure model across each earthquake event.
Although the error varies depending on the model, it remains very small regardless of the ground structure model in the case of a single observation point (from the definition of $ERR$ in Eq.~(\ref{eqerr}), $ERR=0.05$ means that the waveform is matched within an error of about 5\% of its amplitude).
This indicates that the waveform constraints are insufficient when the number of observation points is limited.
In other words, despite the complexity of the time history Green's function in a 3D medium, observed waveforms can be accurately reconstructed regardless of the model used by imposing errors on the estimated input waveform.
Indeed, the estimated input waveform obtained using model000083 (i.e., model number 83 of the 236 candidate models), which exhibits a small error, substantially differs from the true waveform in amplitude and phase characteristics (Fig.~\ref{fig:inputwavenobs1}).
Conversely, increasing the number of observation points becomes imperative to leverage model-specific time-history constraints in Green's functions for accurately reconstructing both observed and input waveforms.
This constraint can be leveraged to diminish the ability to reproduce the observed waveforms depending on the ground structure model, thereby facilitating the selection of the appropriate ground structure model.

\begin{figure}[tb]
\centering
\includegraphics[height=4.2cm]{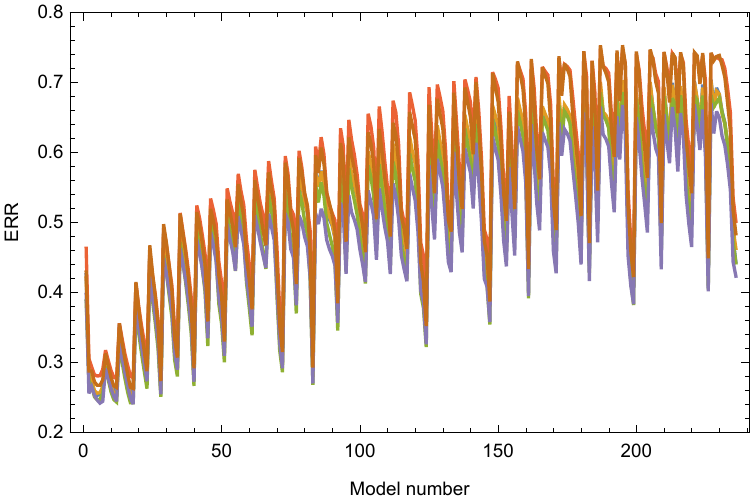}
\includegraphics[height=4.2cm]{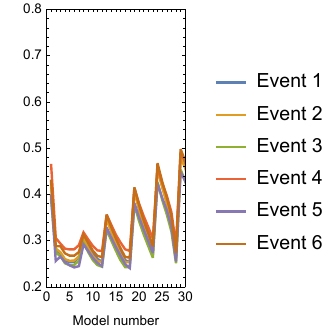}\\
a) $ERR$ for each event (left: overview, right: closeup) \\
\includegraphics[height=4.2cm]{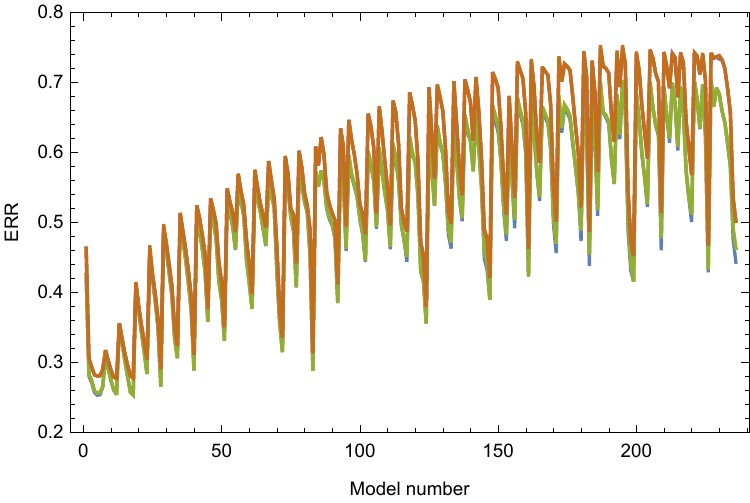}
\includegraphics[height=4.2cm]{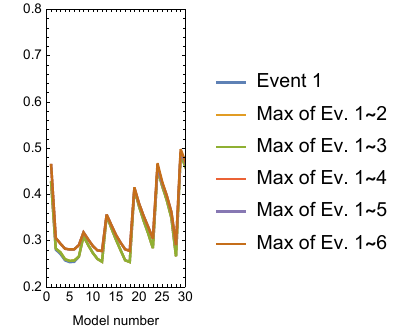}\\
b) Maximum $ERR$ for events (left: overview, right: closeup) \\
\caption{Estimated $ERR$ using nine observation points} 
\label{fig:errornobs9}
\end{figure}

\begin{figure}[tb]
\centering
\includegraphics[height=4.2cm]{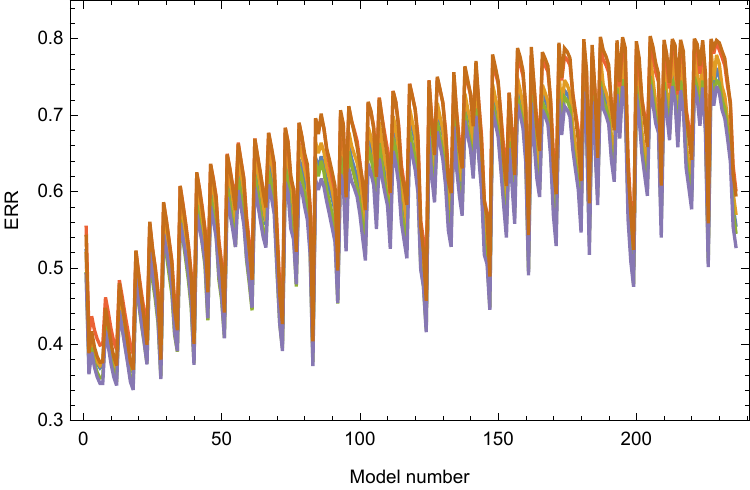}
\includegraphics[height=4.2cm]{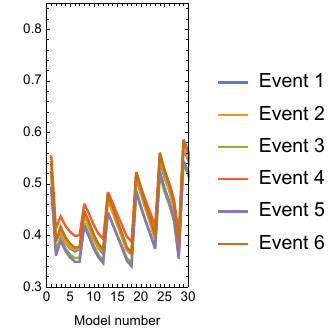}\\
a) $ERR$ for each event (left: overview, right: closeup) \\
\includegraphics[height=4.2cm]{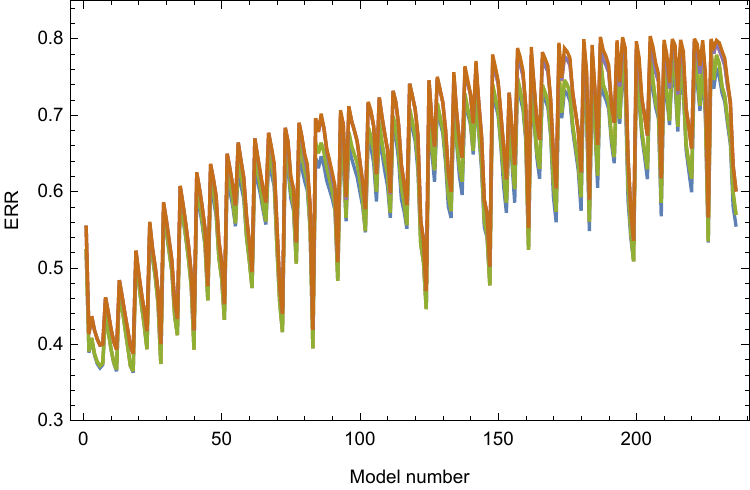}
\includegraphics[height=4.2cm]{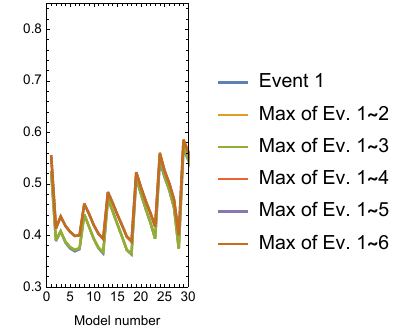}\\
b) Maximum $ERR$ for events (left: overview, right: closeup) \\
\caption{Estimated $ERR$ using 25 observation points} 
\label{fig:errornobs25}
\end{figure}

Next we employ nine observation points on the ground surface $(x_1, x_2)=(100 + 200 i, 100+ 200 j)$ m, where $i,j=0,1,2$.
Figure~\ref{fig:errornobs9}a) illustrates the error obtained for each model across each earthquake event.
Compared with the single observation point case, the information derived from the pseudo-observed earthquake ground motion adequately constrains the ground model.
Consequently, models exhibiting small error levels are consistently identified irrespective of the earthquake event.
The comparison with the historical maximum error depicted in Fig.~\ref{fig:errornobs9}b) reveals that the credible ground model remains consistently selected even as the number of experienced earthquakes and the quantity of information increases.
Furthermore, we augment the information by increasing the number of surface observation points to 25, located at $(x_1, x_2)=(100+100i, 100+100 j)$ m, where $i,j=0,1,....,4$.
The error estimation is shown in Fig.~\ref{fig:errornobs25}a).
As in the 9-point scenario, models with small errors are systematically chosen.
The comparison with the historical maximum error exhibited in Fig.~\ref{fig:errornobs25}b) underscores the continued stable selection of the credible ground model even with the heightened number of experienced earthquakes and augmented information.
Comparing the 9-point case with the 25-point case, it is evident that the increase in information facilitates the stable and systematic extraction of ground models characterized by smaller errors.

\begin{figure}[p]
\centering
\includegraphics[width=0.7\hsize]{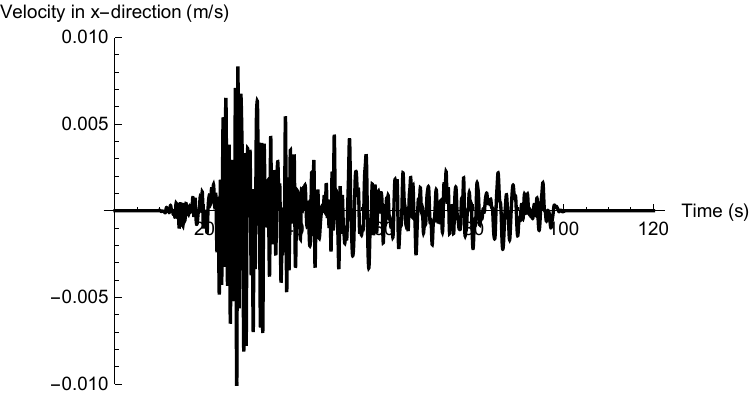}\\
a) True incident wave \\
\includegraphics[width=0.7\hsize]{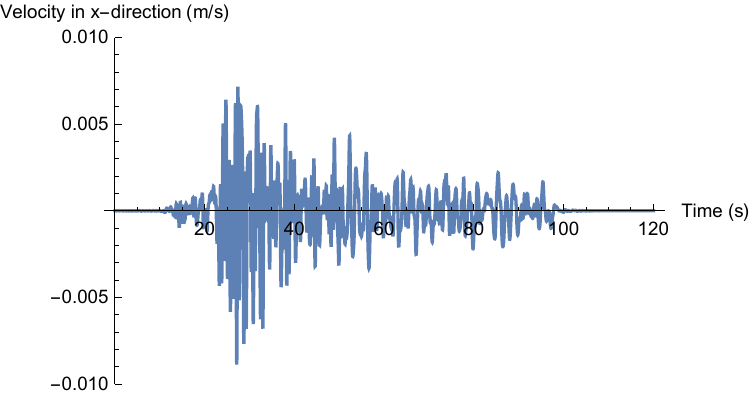}\\
b) Estimated incident wave using model000018 ($ERR=0.388$) \\
\includegraphics[width=0.7\hsize]{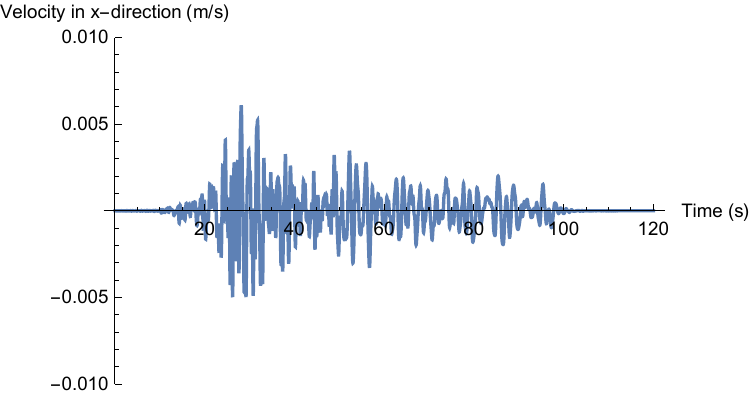}\\
c) Estimated incident wave using model000228 ($ERR=0.781$) \\
\caption{Estimated incident wave using 25 observation points for event \#1. The incident wave can be estimated accurately by selecting a model with a low $ERR$.}
\label{fig:inputwave}
\end{figure}

\begin{figure}[tb]
\centering
\includegraphics[width=\hsize]{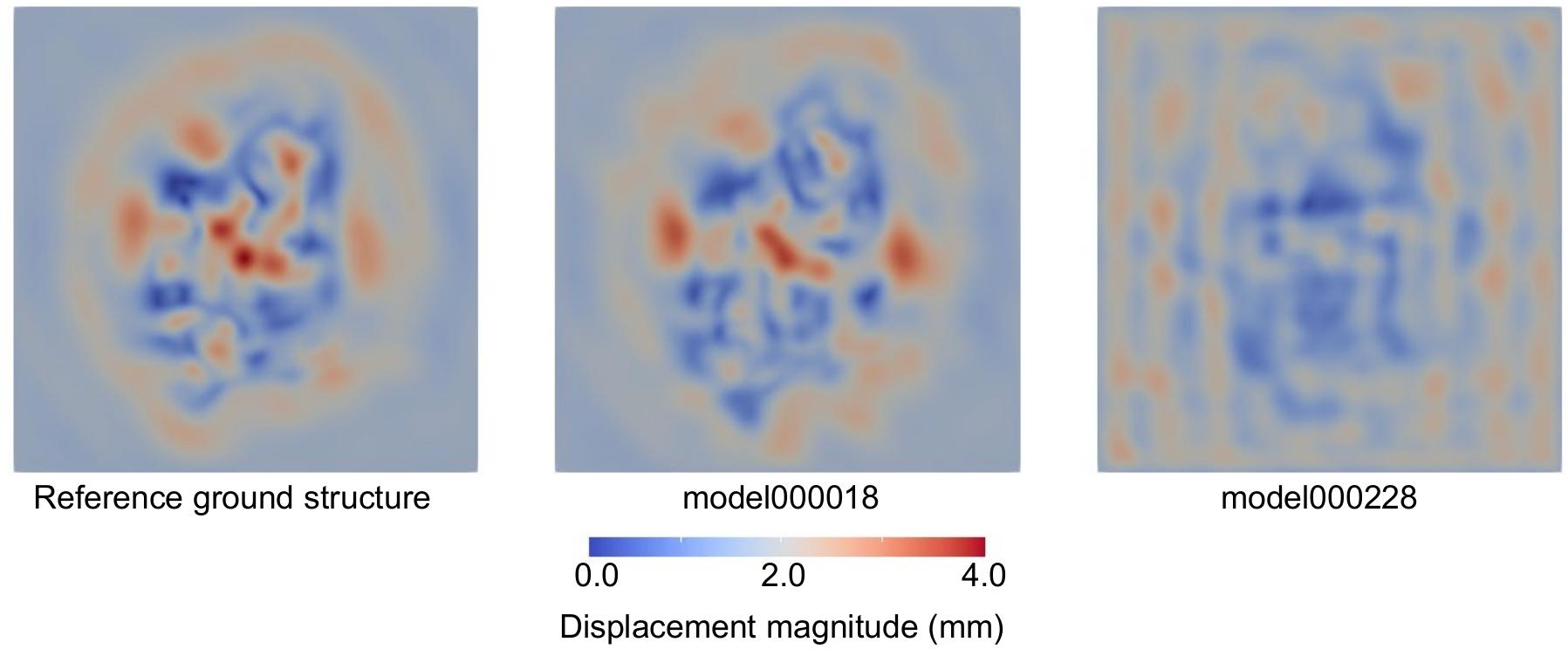}\\
\caption{Displacement response at the surface for the true incident wave (event \#1) at $t=50$ s} 
\label{fig:response}
\end{figure}

\begin{figure}[p]
\centering
\includegraphics[width=0.65\hsize]{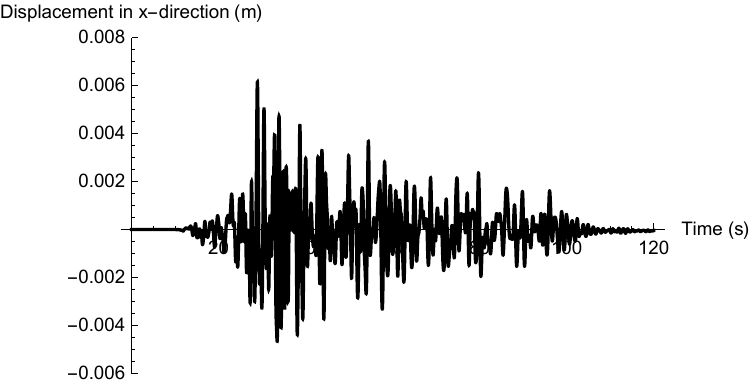}\\
\includegraphics[width=0.65\hsize]{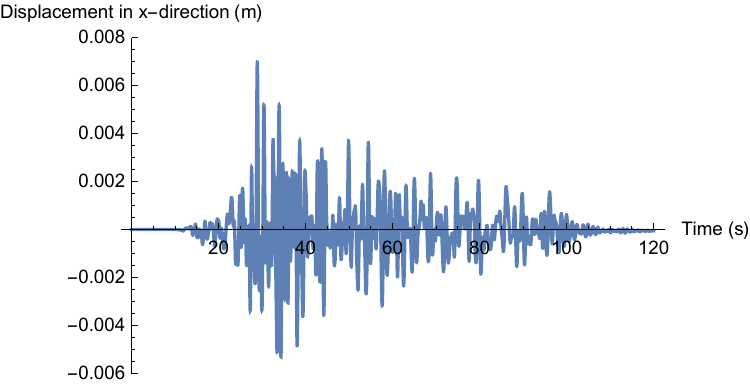}\\
a) Wave at an observation point at $(x_1, x_2)=(300,300)$ m.\\
\includegraphics[width=0.65\hsize]{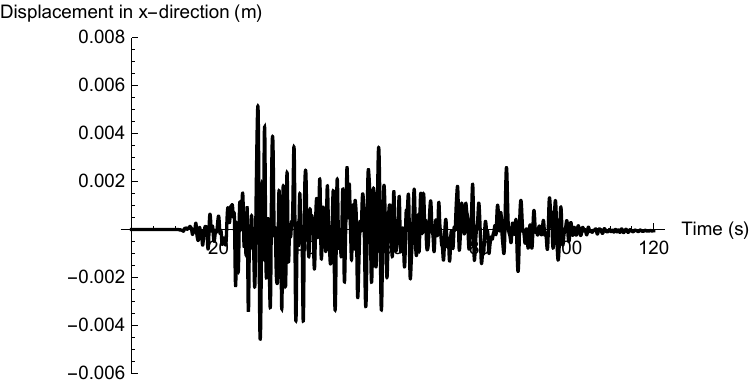}\\
\includegraphics[width=0.65\hsize]{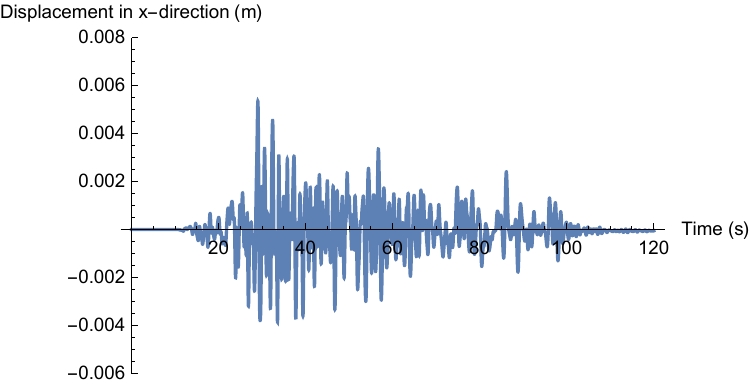}\\
b) Wave at a point on the surface at $(x_1, x_2)=(350,350)$ m, which is far from the observation points. \\
\caption{Displacement time-history response for the true incident wave (event \#1) using the reference model (black) and model000018 (blue).} 
\label{fig:timehistoryresponse}
\end{figure}

Finally, we checked the extracted results.
Figures~\ref{fig:dem}a,b, and c) display the layer thickness distribution of the reference ground model, model000018 (with a minimum error of 0.388 for the case with 25 observation points), and model000228 (with error of 0.781 for the case with 25 observation points).
Model000018, with the smallest error, closely resembles the reference ground model, indicating the extraction of a plausible 3D ground structure model through this method.
While no model among the candidate ground models exactly matches the reference model, model000018 was selected due to its minimal difference compared to the target wavelength used for analysis.
Conversely, model000228, with a large error, substantially deviates from the reference model.
Next, Fig.~\ref{fig:inputwave} illustrates the $x_1$ component of the input seismic motion estimated by model000018 and model000228 using pseudo-observed seismic ground motion at 25 observation points during earthquake event \#1.
The input waveform estimated using model000018, evaluated to have a small error, closely resembles the true input wave, while the waveform estimated using model000228 differs substantially from the true input wave.
This underscores that sufficient observed seismic ground motions enable the Green's function constraint of the time history in the 3D medium obtained from a 3D wavefield analysis, ensuring estimation performance for both observed and input seismic motions when using a plausible model.
Finally, Fig.~\ref{fig:response} depicts the ground motion distribution using the reference model, model000018, and model000228.
The responses of the reference model and model000018 closely align regardless of the location from the observation points, whereas the response of model000228 differs markedly.
As shown in Fig.~\ref{fig:timehistoryresponse}, model000018, selected as the most credible 3D ground structure model in this study, can estimate not only the time series response at the observation point ($(x_1, x_2)=(300, 300)$ m) but also the time series response at a point which is located away from the observation points ($(x_1, x_2)=(350, 350)$ m).
This indicates that a consistent ground model that matches the observed seismic ground motion has been extracted.
These results underscore the significance of selecting a plausible 3D ground structure model, as demonstrated in this study, for mitigating earthquake damage, since even 3D ground structure models estimated by geotechnical engineering methods exhibit substantial differences in their performance in reproducing seismic ground motions.

As described above, it is evident that a 3D ground structure model demonstrating high performance in reproducing observed seismic motions can be extracted by using seismic motion observations at the ground surface.
Achieving this necessitates multiple 3D seismic response analyses.
In this study, finite element models are generated for the 236 ground models with varying layer thicknesses, and for each model, the Green's function response of Eq.~(\ref{GE:ORG}) to the unit impulse wave $p(t)$ in each of the $x_1$, $x_2$, and $x_3$ directions are computed, resulting in a total of 236$\times$3 = 708 instances of 3D wave field analysis.
An automatic method for generating high-quality finite element models is required to facilitate such a large number of analyses.
Moreover, a fast implicit time integration-based 3D seismic analysis method capable of performing stable calculations, even on finite element models with small local elements that are essential to faithfully model the complex ground geometry, is required.
Consequently, in this study, finite element models comprising second-order tetrahedral elements are automatically generated using a mesh generator \cite{meshgenerator} employing octree background cells (the finite element models utilized in this study average 2,320,401 degrees of freedom).
The Newmark-$\beta$ method, a type of implicit time integration, is employed to compute the seismic ground motion over 16,000 time steps with a time increment of $dt=0.001$ s to ensure stability.
While implicit time integration is suitable for stable computation, it entails solving solutions of large sparse matrix equations, resulting in substantial computational costs.
To address this issue, we use a fast solver from \cite{Kusakabe2022EESD}, based on the conjugate gradient method with variable preconditioning \cite{SC14}.
Consequently, even a 3D seismic analysis with 2.32 million degrees of freedom (516748 second-order tetrahedral elements with minimum element size of 5 m) and 16,000 time steps can be executed in approximately 270 s on a computing environment with an NVIDIA A100 PCIe 40 GB GPU \cite{A100}.
The entire evaluation can be completed stably within a short duration of approximately 1.5 h using 44 A100 GPUs.
Due to time constraints, conducting 3D wave propagation analysis for numerous ground models of this scale would have been impractical in the past. However, leveraging a fast 3D wave propagation method, as demonstrated here, enables such analyses to be conducted within feasible timeframes.

\section{Concluding Remarks}

This study introduced a method to extract a 3D ground structure model that can reproduce observed seismic ground motions from a pool of candidate ground structure models using surface-observed earthquake ground motions and 3D seismic ground motion analysis.
Through numerical experiments, we illustrated its effectiveness.
Even when ground models are generated using methods based on geotechnical engineering aspects, their performance in reproducing observed ground motion varies substantially.
A comparison of simulation results between models exhibiting high and low performance underscored substantial differences in seismic damage estimation during earthquakes, highlighting the efficacy of our method.
While our numerical experiments focused solely on treating the layer boundary geometry as a model parameter, there are no constraints on parameterization in this context.
Both the material properties and geometry can serve as model parameters for extracting a plausible 3D ground structure model.
The above findings indicate that our proposed method is expected to contribute to enhancing the reliability of 3D ground structure models generated by supplementing borehole data and other databases, which is expected to become more common in the future.
There is also possibility in enhancing the method by utilizing further HPC and AI for efficiently generating better candidate models around the ground models that are chosen to be credible and searching for a ground model with better performance in reproducing observed ground motions.

\subsection*{Acknowledgments} 

This work used waveforms provided by NIED KiK-net, National Research Institute for Earth Science and Disaster Resilience, Japan.

\end{document}